\documentclass[showpacs,aps,prd,nofootinbib,floatfix,amsmath,amssymb]{revtex4}
\usepackage{graphicx}
\begin{document}

\makeatletter
\newbox\slashbox \setbox\slashbox=\hbox{$/$}
\newbox\Slashbox \setbox\Slashbox=\hbox{\large$/$}
\def\pFMslash#1{\setbox\@tempboxa=\hbox{$#1$}
  \@tempdima=0.5\wd\slashbox \advance\@tempdima 0.5\wd\@tempboxa
  \copy\slashbox \kern-\@tempdima \box\@tempboxa}
\def\pFMSlash#1{\setbox\@tempboxa=\hbox{$#1$}
  \@tempdima=0.5\wd\Slashbox \advance\@tempdima 0.5\wd\@tempboxa
  \copy\Slashbox \kern-\@tempdima \box\@tempboxa}
\def\FMslash{\protect\pFMslash}
\def\FMSlash{\protect\pFMSlash}
\def\miss#1{\ifmmode{/\mkern-11mu #1}\else{${/\mkern-11mu #1}$}\fi}
\makeatother

\title{Integration of Kaluza--Klein modes in Yang--Mills theories}
\author{H. Novales--S\' anchez and J. J. Toscano}
\address{Facultad de Ciencias F\'{\i}sico Matem\'aticas,
Benem\'erita Universidad Aut\'onoma de Puebla, Apartado Postal
1152, Puebla, Puebla, M\'exico.}
\begin{abstract}
A five dimensional pure Yang--Mills theory, with the fifth coordinate compactified on the orbifold $S^1/Z_2$ of radius $R$, leads to a four dimensional theory which is governed by two types of infinitesimal gauge transformations, namely, the well known standard gauge transformations (SGT) dictated by the ${\rm SU}_4(N)$ group
under which the zero Fourier modes $A^{(0)a}_\mu$ transform as gauge fields, and a set of nonstandard gauge transformations (NSGT) determining the gauge nature of the Kaluza--Klein (KK) excitations $A^{(m)a}_\mu$. By using a SGT--covariant gauge--fixing procedure for removing the degeneration associated with the NSGT, we integrate out the KK excitations and obtain a low--energy effective Lagrangian expansion involving all of the independent canonical--dimension--six operators that are invariant under the SGT of the ${\rm SU}_4(N)$ group and that are constituted by light gauge fields, $A^{(0)a}_\mu$, exclusively. It is shown that this effective Lagrangian is invariant under the SGT, but it depends on the gauge--fixing of the gauge KK excitations. Our result shows explicitly that the one--loop contributions of the KK excitations to light (standard) Green's functions are renormalizable.
\end{abstract}

\pacs{11.10.Kk, 11.15.-q, 14.70.Pw, 14.80.Rt}

\maketitle

\section{Introduction}
\label{int}
Extra dimensional Standard Model (SM) extensions incarnate attractive proposals to explore physics governing nature at scales not reached by collider experiments so far. SM extensions involving universal extra dimensions~\cite{UED1} (UED) are particularly alluring as they set a relatively small lower bound on the compactification scale, at 300GeV. A key issue of this type of extra dimensional models is that hitherto they are the only ones whose KK gauge sector has been~\cite{NT} consistently quantized, which is crucial to perform phenomenological calculations as all of the low--energy KK contributions generated by models with UED involve quantum effects, so that a precise knowledge of the complete KK ghost sector is imperious. These models invoke at least two main high energy scales, namely, that corresponding to the size of the extra dimension and another more fundamental, denoted by $M_{\rm S}$, which could be perhaps the strings scale. So, UED extensions of the SM are not fundamental theories, but low energy manifestations of the underlying theory beyond the fundamental scale.

In this paper, our work context shall be an ${\rm SU}_5(N)$--invariant Yang--Mills theory defined in a five dimensional space--time manifold in which the extra dimension shall be supposed to be universal and shall be compactified on the orbifold $S^1/Z_2$, with radius $R$. The ordinary four dimensional coordinates shall be denoted by $x$, while the extra dimension shall be labeled by $y$. Five dimensional Lorentz indices shall be denoted by capital roman letters ($M,N,\ldots$), four dimensional Lorents indices shall be represented by greek letters ($\mu,\nu,\ldots$), gauge group indices shall be denoted by lower--case roman letters ($a,b,\ldots$), and KK mode indices shall be always placed between parentheses. We consider the five dimensional Yang--Mills Lagrangian,
\begin{equation}
\label{5DYM}
{\cal L}_{\rm 5YM}=-\frac{1}{4}{\cal F}^a_{MN}{\cal F}^{aMN},
\end{equation}
with the curvature ${\cal F}^a_{MN}$ defined as usual:
\begin{equation}
{\cal F}^a_{MN}=\partial_M{\cal A}^a_N-\partial_N{\cal A}^a_M+g_5f^{abc}{\cal A}^b_M{\cal A}^c_N ,
\end{equation}
where ${\cal A}^a_M$ represents the five--dimensional gauge fields, $g_5$ stands for the five--dimensional coupling constant, which has dimensions of $({\rm mass})^{-1/2}$, and the $f^{abc}$ are the structure constants. The fact that the coupling constant $g_5$ is dimensionfull indicates that this theory is non--renormalizable. As mentioned above, there is an  underlying theory at a scale $M_{\rm S}$, beyond the compactification scale, and the effects of such physics can be, in principle, parametrized by means of an effective Lagrangian expansion consisting in the five--dimensional Yang--Mills theory and an infinite sum of higher--than--five canonical dimension operators as
\begin{equation}
{\cal L}^{\rm eff}_{\rm 5D}={\cal L}_{\rm 5YM}({\cal A}_M^a)+\sum_k\beta_k \frac{g_5^{n_k}}{M_S^{r_k}}\hspace{0.05cm}{\cal O}_k^{\rm 5D}({\cal A}^a_M).
\end{equation}
The ${\cal O}^{\rm 5D}_k$ are operators of canonical dimension higher than five that are constituted, exclusively, by the five dimensional gauge fields of the extra dimensional Yang--Mills theory, as well as the covariant derivative. The $\beta_k$ coefficients are dimensionless quantities that parametrize, in a model--independent manner, the effects of the fundamental theory that describes the physics at the $M_{\rm S}$ scale. Each of the effective operators has a coefficient that involves appropriate powers of the coupling constant and the $M_{\rm S}$ energy scale, so that each term has the correct dimension of mass. As the extra dimensional theory is non--renormalizable in the Dyson's sense, there is no criterion that restricts the number of effective operators to include in the series. One can compactify the fifth dimension and integrate it in the action to obtain an effective four dimensional theory in which the dynamic variables are the KK modes. It is worth emphasizing that, after compactification, we KK--expand the five dimensional curvatures, which are the covariant objects of the theory, as such a procedure ensures the preservation of gauge invariance at the four dimensional level and leads~\cite{NT,NTbr} to a {\it pure--gauge KK theory} (GKKT), which is~\cite{NT} separately invariant under two sorts of infinitesimal gauge transformations: the {\it standard gauge transformations} (SGT), under which the zero KK modes are gauge fields; and the {\it nonstandard gauge transformations} (NSGT), that transform the KK excited modes as gauge fields. After compactification and integration of the extra dimension, one obtains a theory of the form
\begin{equation}
{\cal L}^{\rm eff}_{\rm 4D}={\cal L}_{\rm 4YM}(A^{(0)a}_\mu,A^{(m)a}_\mu)+\sum_{k>4}\frac{\alpha_k}{M_{\rm S}^{k-4}}\hspace{0.05cm}{\cal O}_k^{\rm 4D}(A^{(0)a}_\mu,A^{(m)a}_\mu),
\label{L5Deff}
\end{equation}
with $A^{(0)a}_\mu$ and $A^{(m)a}_\mu$ representing the gauge KK zero and excited modes, respectively, and where
\begin{equation}
{\cal L}_{\rm 4YM}=\int_0^{2\pi R} dy\hspace{0.1cm}{\cal L}_{\rm 5YM}.
\end{equation}
The ${\cal L}_{\rm 4YM}$ Lagrangian has interesting features that deserve to be mentioned. This Lagrangian is invariant under the SGT and the NSGT, and its structure involves KK curvatures that vary covariantly under both sets of gauge transformations. Such curvatures depend on the gauge KK modes and also contain the pseudo--Goldstone bosons, $A^{(m)a}_5$, which can be removed~\cite{NT} from the theory by an appropriate fixation of the gauge. As commented above, the extra dimensional theory is non--renormalizable, but it is interesting noting that the four dimensional  coupling constant of the ${\cal L}_{\rm 4YM}$ GKKT is dimensionless and that the canonical dimension of the couplings of this Lagrangian is equal or less than four, as required by Dyson's criterion. The non--renormalizable nature of the five dimensional theory manifests itself at the four dimensional level through the infinite sums over the KK modes, which introduce a divergent behavior. A striking quality of the ${\cal L}_{\rm 4YM}$ gauge KK Lagrangian is that it is~\cite{NT} renormalizable at the one--loop level, which does not hold at higher orders or when two or more extra dimensions are considered. This is consistent with the fact that in UED models with only one extra dimension the KK sums are~\cite{UED1,UED2} convergent, which is related to the nonsensitivity of the low--energy observables corrections with respect to the cut--off $M_{\rm S}$. This property is very important and phenomenological examples do exist~\cite{FMNRT}. On the other hand, the ${\cal O}^{\rm 4D}_k$, in Eq.(\ref{L5Deff}), are combinations of effective operators that have canonical dimension higher than four and that are composed by the gauge KK modes. Moreover, these objects are invariant under the four dimensional Lorentz transformations as well as under the SGT and the NSGT. The lowest--order corrections to light Green's functions that can be generated by the ${\cal L}_{\rm 4YM}$ Lagrangian are radiative corrections at the one--loop level. Such a property comes from the KK parity conservation~\cite{UED1} that occurs in the context of UED. Contrastingly, the effective operators ${\cal O}^{\rm 4D}_k$ produce tree--level corrections to low--energy Green's functions. The most important contributions of these operators are produced by the vertices involving exclusively zero KK modes, which are the lightest fields. However, the supression $1/M_{\rm S}^k$ on the effective operators renders~\cite{FMNRT} them dominated by the one--loop effects of the ${\cal L}_{\rm 4YM}$ Lagrangian, as expected~\cite{UED1,UED2} for UED models.

The appropriate quantization of the ${\cal L}_{\rm 4YM}$ Lagrangian requires the fixation of the gauge, which comes along with the derivation~\cite{NT} of the most general KK ghost sector through the Becchi--Rouet--Stora--Tyutin (BRST) formalism~\cite{BRST}. The gauge--fixing procedure can be accomplished~\cite{NT} in such a way that the invariance with respect to the NSGT is removed, but that with respect to the SGT is still fulfilled. This can be achieved by utilizing a gauge--fixing scheme that is similar to another one proposed~\cite{331tM} some years ago in the context of the so--called 331 model~\cite{331}. Within this approach, a set of SGT--covariant gauge--fixing functions is introduced~\cite{NT,FMNRT} and the theory can be quantized with respect to the KK excited modes, but leaving the zero modes as classical fields. The resulting quantum Lagrangian, which remains invariant under the SGT, is constituted by three parts: the ${\cal L}_{\rm 4YM}$ Lagrangian, which we defined above; the gauge--fixing  term, which is compounded by the SGT--covariant gauge--fixing functions; and the Faddeev--Popov ghost part, which naturally emerges from the BRST formalism, and which also depends on the gauge--fixing functions. The main concern of the present work is the functional integration of the KK excited modes in the quantum Lagrangian, which are the heavy fields of the KK theory, and the derivation of an effective Lagrangian low--energy expansion that involves only zero KK modes, which are the light fields of the theory. The SGT--covariant gauge--fixing approach~\cite{NT} possesses advantages from the practical viewpoint, and in this work we take advantage of them. In fact, we prove that such gauge--fixing procedure renders the contributions of the ghost fields minus twice those of the pseudo--Goldstone bosons. As the heavy fields to integrate out are gauge fields, some interesting issues arise. In order to perform the KK heavy modes integration, the gauge--fixing and the Faddeev--Popov ghost terms must be taken into account. The ghost term depends on the gauge--fixing functions, which means that both the gauge--fixing and the Faddeev--Popov Lagrangians involve the gauge--fixing parameter, $\xi$. This observation is very important, as such dependence on the gauge--fixing parameter is inherited by the low--energy effective theory,
\begin{equation}
{\cal L}_{\rm eff}^\xi={\cal L}_{\rm YM}(A^{(0)a}_\mu)+\sum_m\sum_{k>4}\frac{\varepsilon_k(\xi)}{(\frac{m}{R})^k}{\cal O}_k(A^{(0)a}_\mu;\xi)+\sum_{k>4}\frac{\alpha_k}{M_{\rm S}^{k-4}}{\cal O}^{\rm 4D}_k(A^{(0)a}_\mu).
\label{eff4DLgeneralformxi}
\end{equation}
In this effective Lagrangian, we have considered the low--energy four dimensional Yang--Mills ${\rm SU}_4(N)$--invariant theory, described by the ${\cal L}_{\rm YM}$ Lagrangian. The second term of the effective theory is a sum of non--renormalizable higher--than--four canonical dimension operators that carry the one--loop effects of the KK excited modes on low--energy Green's functions. Finally, the third term represents a sum of the non--renormalizable operators originated, at the five dimensional level, in the higher--than--five canonical dimension operators (see Eq.(\ref{L5Deff})) after compactifaying the fifth dimension and KK--expanding the five dimensional covariant objects. In this effective theory, we have disregarded all of the contributions produced by KK excited modes that do not impact light Green's functions at the one--loop level.  We have also ignored, in the third term, all of the non--renormalizable terms that involve KK excited modes, for they comprehend the most suppressed effects. The coefficients $\varepsilon_k(\xi)$ are gauge--dependent dimensionless parameters that quantify the effects produced, at the one--loop level, by the excited KK modes on light Green's functions at an energy scale that is small when compared with the compactification scale, $R^{-1}$. The ${\cal O}_k$ are gauge dependent linear combinations of operators of canonical dimension higher than four. They are governed by the low--energy symmetries, which are the invariance with respect to both the four dimensional Lorentz transformations and the SGT. Note that the first series of non--renormalizable operators in Eq.(\ref{eff4DLgeneralformxi}) involves the mass of the gauge KK excited modes, which is given, for the $m$--th mode, by $m_m=m/R$, where $m$ is an integer number. There is a sum running over all of the KK modes, so that for each term of the sum over $k$ there is a Riemman $\zeta$--function, which is finite. This fact explicitly demonstrates that the KK--sums of excited modes contributions to light Green's functions are convergent, as expected~\cite{NT}. In this paper, we integrate out the KK excited modes generated by the five dimensional Yang--Mills Lagrangian and obtain the explicit expressions of the terms that involve the canonical--dimension--six operators that are invariant under the SGT, and find that they do not involve UV divergencies. This result shows that the one--loop contributions of the KK excited modes on light Green's functions are renormalizable, as it was recently proven in Ref.\cite{NT} and phenomenologically illustrated in Ref.\cite{FMNRT}. As mentioned above, these results are gauge dependent, as they contain the gauge--fixing parameter. The integration of heavy gauge fields, accomplished in the present paper, is a novel calculation. In fact, to our knowledge, there is no work so far concerning the integration of heavy gauge modes to obtain an effective low--energy expansion, and the possibility of a gauge dependent effective Lagrangian has not been pointed out. To achieve the KK excited modes integration, we adjust the method proposed in Ref.\cite{BS} so that it works in the case of massive gauge fields, and derive a low--energy effective Lagrangian expansion comprehending up to canonical--dimension--six non--renormalizable operators that are subjected to the low--energy symmetries.

The paper is organized as follows. In Section \ref{def} we outline the procedure to obtain the GKKT theory from the five dimensional Yang--Mills Lagrangian, Eq.(\ref{5DYM}), and briefly discuss some issues concerning the gauge structure of such four dimensional theory and the structure of its quantum version. Section \ref{KKinte} is dedicated to integrate out the KK excited modes, first within the Feynman--'t Hooft context and then in the general $R_\xi$ gauge. Finally, in Appendix \ref{appA}, we provide a the details of the derivation of a low--energy expansion when heavy gauge fields are integrated out.

\section{The four dimensional pure--gauge Kaluza--Klein theory}
\label{def}
In this section, we obtain a four dimensional KK theory from the five dimensional ${\rm SU}_5(N)$--invariant Yang--Mills Lagrangian, Eq(\ref{5DYM}), by compactifying the extra dimension, which is supposed to be universal, in the orbifold $S^1/Z_2$ and integrating it in the action,
\begin{equation}
{\cal S}=\int d^4x\int_0^{2\pi R} dy\hspace{0.05cm}{\cal L}_{\rm 5YM}\equiv\int d^4x\hspace{0.05cm}{\cal L}_{\rm 4YM}.
\end{equation}
Within the BRST formalism~\cite{BRST}, the gauge parameters defining the five dimensional gauge transformations coincide with the ghost fields. This means that in the case of UED models such parameters also propagate in the fifth dimension, so that they can be KK--expanded. The whole set of gauge parameters KK modes defines an infinite set of local gauge transformations, which can be separated into the SGT and the NSGT. In order to preserve enough gauge invariance when passing from five to four dimensions, one must KK-expand covariant objects, which, in the case of Eq.(\ref{5DYM}), are the five dimensional curvatures. This approach leads~\cite{NT} to the four dimensional GKKT
\begin{equation}
{\cal L}_{\rm 4YM}=-\frac{1}{4}\left( {\cal F}^{(0)a}_{\mu\nu}{\cal F}^{(0)a\mu\nu}+{\cal F}^{(m)a}_{\mu\nu}{\cal F}^{(m)a\mu\nu}+2{\cal F}^{(m)a}_{\mu5}{\cal F}^{(m)a\mu5} \right),
\label{L4YM}
\end{equation}
with the four dimensional curvatures given by~\cite{NT}
\begin{eqnarray}
{\cal F}^{(0)a}_{\mu\nu}&=&F^a_{\mu\nu}+gf^{abc}A^{(m)b}_\mu A^{(m)c}_\nu,
\label{curv0}
\\ \nonumber \\
{\cal F}_{\mu\nu}^{(m)a}&=&{\cal D}^{ab}_\mu A^{(m)b}_\nu-{\cal D}^{ab}_\nu A^{(m)b}_\mu+gf^{abc}\Delta^{mrn}A^{(r)b}_\mu A^{(n)c}_\nu,
\label{curvgm}
\\ \nonumber \\
{\cal F}^{(m)a}_{\mu5}&=&{\cal D}^{ab}_\mu A^{(m)b}_5+\frac{m}{R}A^{(m)a}_\mu+gf^{abc}\Delta'^{mrn}A^{(r)b}_\mu A^{(n)c}_5,
\label{curvGm}
\end{eqnarray}
where ${\cal D}^{ab}_\mu$ is the covariant derivative in the adjoint representation of the ${\rm SU}_4(N)$ group. In the last expressions, any pair of repeated indices, included the modes ones, denotes a sum. The $A^{(m)a}_5$ fields are the KK modes of the fifth component of the five dimensional vector bosons. The factors $\Delta^{mrn}$ and $\Delta'^{mrn}$ are linear combinations of Kroenecker deltas, and their specific forms are irrelevant for the present work. The four dimensional coupling constant is denoted, as usual, by $g$. The zero--mode curvature, ${\cal F}^{(0)a}_{\mu\nu}$, involves the ordinary four dimensional Yang--Mills curvature,
\begin{equation}
F^a_{\mu\nu}=\partial_\mu A^{(0)a}_\nu-\partial_\nu A^{(0)a}_\mu+gf^{abc}A^{(0)b}_\mu A^{(0)c}_\nu ,
\end{equation}
which is made only of zero--mode gauge fields. With this in mind, notice that the ${\cal L}_{\rm 4YM}$ Lagrangian contains the ordinary four dimensional ${\rm SU}_4(N)$--invariant Yang--Mills theory, which is exclusively constituted by light fields, that is, by KK zero--mode fields. The ${\cal L}_{\rm 4YM}$ Lagrangian is invariant under the SGT
\begin{eqnarray}
\delta A^{(0)a}_\mu&=&{\cal D}^{ab}_\mu\alpha^{(0)b},
\\ \nonumber \\
\delta A^{(m)a}_\mu&=&gf^{abc}A^{(m)b}_\mu\alpha^{(0)c},
\\ \nonumber \\
\delta A^{(m)a}_5&=&gf^{abc}A^{(m)b}_5\alpha^{(0)c},
\end{eqnarray}
which are clearly defined by the zero--mode gauge parameters, $\alpha^{(0)a}$. Note that, under such gauge variations, the zero--modes $A^{(0)a}_\mu$ transform standardly as gauge fields, while the KK excitations $A^{(m)a}_\mu$ and $A^{(m)a}_5$ behave as matter fields transforming in the adjoint representation of the ${\rm SU}_4(N)$ gauge group. On the other hand, the ${\cal L}_{\rm 4YM}$ Lagrangian is also invariant under following set of NSGT,
\begin{eqnarray}
\delta A^{(0)a}_\mu&=&gf^{abc}A^{(n)b}_\mu\alpha^{(n)c},
\\ \nonumber \\
\delta A^{(m)a}_\mu&=&{\cal D}^{(mn)ab}_\mu\alpha^{(n)b},
\\ \nonumber \\
\delta A^{(m)a}_5&=&{\cal D}^{(mn)ab}_5\alpha^{(n)b},
\end{eqnarray}
which, contrastingly to the case of the SGT, are determined exclusively by the KK excitations of the gauge parameters. The nature of the KK modes with respect to the NSGT is appreciably different in comparison with their comportment under the SGT. The zero modes $A^{(0)a}_\mu$ are not gauge fields under the NSGT, but they transform in a way that resembles the adjoint transformation, although this variation involves a mixing among KK excitations of the gauge parameters and those of the vector bosons. The NSGT of the KK excited modes $A^{(m)a}_\mu$ contain the object
\begin{equation}
{\cal D}^{(mn)ab}_\mu=\delta^{mn}{\cal D}^{ab}_\mu-gf^{abc}\Delta^{mrn}A^{(r)c}_\mu,
\end{equation}
which is a sort of covariant derivative. Thus these KK excitations transform as gauge fields with respect to the NSGT. Finally, the object
\begin{equation}
{\cal D}_5^{(mn)ab}=-\delta^{mn}\delta^{ab}\frac{m}{R}-gf^{abc}\Delta'^{mrn}A^{(r)c}_5
\end{equation}
is present in the NSGT of the KK scalar excitations $A^{(m)a}_5$. The scalar fields $A^{(m)a}_5$ can be completely removed from the theory  by performing the particular NSGT
\begin{equation}
\delta A^{(m)a}_5=\frac{R}{m}{\cal D}^{(mn)ab}_5 A^{(m)a}_5,
\end{equation}
for which the KK excited gauge parameters have been taken to be $\alpha^{(m)a}=(R/m)A^{(m)a}_5$. This result shows explicitly that such scalars are pseudo--Goldstone bosons whose degrees of freedom have been eaten by the KK excited gauge modes $A^{(m)a}_\mu$ to aquire their masses. It is worth mentioning that the KK curvatures ${\cal F}^{(0)a}_{\mu\nu}$, ${\cal F}^{(m)a}_{\mu\nu}$ and ${\cal F}^{(m)a}_{\mu5}$, exhibited in Eqs.(\ref{curv0}), (\ref{curvgm}) and (\ref{curvGm}), transform covariantly under the SGT and the NSGT, which is crucial to elegantly prove that the ${\cal L}_{\rm 4YM}$ Lagrangian is invariant under both sets of gauge transformations.

The quantization of this GKKT can be consistently executed~\cite{NT} on the grounds of the BRST formalism~\cite{BRST}. A remarkable attribute of this theory is the possibility of splitting the quantization procedure into two independent parts. One can, for instance, fix the gauge with respect to the NSGT and then quantize the KK excited modes, while leaving the gauge invariance with respect to the SGT and preserving the gauge zero modes as classical background gauge fields. For the present work, we follow this route. However, it is worth emphasizing that one can, if desired, quantize the zero modes as usual or in an unconventional manner. Within the BRST approach, the gauge--fixing, ${\cal L}_{\rm GF}$, and Faddeev--Popov, ${\cal L}_{\rm FPG}$, terms are derived. Both sectors depend crucially on a set of gauge--fixing functions, denoted by $f^{(m)a}$, which can be suitably chosen so that the quantum Lagrangian remains invariant under the SGT, although there is no more degeneracy with respect to the NSGT. A convenient election of the gauge--fixing functions is the following SGT--covariant set:
\begin{equation}
f^{(m)a}={\cal D}^{ab}_\mu A^{(m)b\mu}-\xi\frac{m}{R}A^{(m)a}_5.
\label{gffunc}
\end{equation}
Recall that $\xi$ is the gauge--fixing parameter. With this choice, the quantum Lagrangian is found to be
\begin{eqnarray}
{\cal L}_{\cal Q}&=&{\cal L}_{\rm 4YM}+{\cal L}_{\rm GF}+{\cal L}_{\rm FPG},
\label{Lquantum}
\end{eqnarray}
with ${\cal L}_{\rm 4YM}$ given by Eq.(\ref{L4YM}). The gauge--fixing term is explicitly given by
\begin{eqnarray}
{\cal L}_{\rm GF}&=&-\frac{1}{2\xi}({\cal D}^{ab}_\mu A^{(m)b\mu})({\cal D}^{ac}_\nu A^{(m)c\nu})+m_m A^{(m)a}_5({\cal D}^{ab}_\mu A^{(m)b\mu})-\frac{1}{2}\xi m_m^2A^{(m)a}_5A^{(m)a}_5,
\label{Lgf}
\end{eqnarray}
while the Faddeev--Popov Lagrangian can be divided into two parts as
\begin{equation}
{\cal L}_{\rm FPG}={\cal L}_{\rm FPG}^1+{\cal L}_{\rm FPG}^2,
\end{equation}
where
\begin{eqnarray}
{\cal L}_{\rm FPG}^1&=&\bar{C}^{(m)b}({\cal D}^{ba}_\mu\hspace{0.05cm}{\cal D}^{ac\mu})C^{(m)c}-\xi m_m^2\bar{C}^{(m)a}C^{(m)a}-gf^{abc}\left[ \Delta^{mrn}\bar{C}^{(m)d}({\cal D}^{ad}_\mu A^{(r)c\mu})C^{(n)b}
\right.
\nonumber \\ \nonumber \\
&&-\frac{1}{\xi}\Delta^{mrn}\bar{C}^{(r)c}({\cal D}^{ad}_\mu A^{(m)d\mu})C^{(n)b}
\left.
+\xi m_m\Delta'^{mrn} \bar{C}^{(m)a}A^{(r)c}_5C^{(n)b}-m_m\Delta^{mrn}\bar{C}^{(r)a}A^{(m)c}_5C^{(n)b}
\right].
\label{Lfpg}
\end{eqnarray}
Here, $C^{(m)a}$ ($\bar{C}^{(m)a}$) stands for the KK ghost (antighost) fields excitations. The ${\cal L}_{\rm FPG}^2$ term is constituted~\cite{NT} by quartic interactions among KK ghost fields, and its specific structure is not needed to achieve the purposes pursued in the present paper, so that we shall omit it from here on. A notable quality of the gauge--fixing functions given in Eq.(\ref{gffunc}) is that they lead to the cancelation of the non--physical bilinear and trilinear couplings $A^{(m)a}_\mu A^{(n)b}_5$ and $A^{(0)a}_\mu A^{(m)b}_\nu A^{(n)c}_5$. This issue shall be important in the next section, when we integrate out the heavy KK modes.

Extra dimensional models involve dimensionfull coupling constants, which in turn implies that they are non--renormalizable, and this, of course, also holds for the five dimensional ${\rm SU}_5(N)$--invariant Yang--Mills Lagrangian that we took as our starting point. Nonetheless, by examining the structure of each term of the ${\cal L}_{\cal Q}$ Lagrangian, Eqs.(\ref{L4YM}), (\ref{Lgf}) and (\ref{Lfpg}), one can perceive that the four dimensional coupling constant is dimensionless and that all of the couplings have canonical dimension equal or less than four, as required by Dyson's renormalizability criterion. In general, KK theories involve infinite sums over the KK modes that must be also performed when calculating corrections to light Green's functions. The non--renormalizability of a given extra dimensional theory reveals itself at the four dimensional level through these infinite sums. In other words, the divergencies present at the extra dimensional level persist in four dimensions and are produced by discrete rather than continuous sums. Particularly, theories with only one extra dimension do not introduce this sort of divergencies at the four dimensional level when inserted into one--loop corrections to low--energy observables. This property is not fulfilled when two or more extra dimensions are considered, or in the case of two--loops and beyond corrections. The quantum Lagrangian ${\cal L}_{\cal Q}$ produces~\cite{NT} renormalizable one--loop contributions to light Green's functions, as all of the divergencies introduced by the KK excited modes are absorbed by the parameters of the light theory. This feature is very important because the fact that the divergencies can be controlled ensures that this sort of quantum corrections lead to unambiguous results. In this context, the one--loop KK corrections to the light Green's functions $WW\gamma$ and $WWZ$ have been calculated~\cite{FMNRT}, finding gauge--dependent, although well--behaved results that are nonsensitive to the cut--off $M_{\rm S}$.

\section{Integration of the heavy KK modes}
\label{KKinte}
In this section, we integrate out the KK excited modes of ${\cal L}_{\cal Q}$, which are the heavy fields of the theory, and obtain a low--energy effective Lagrangian that depends only on the light fields, personified by the KK zero modes. The obtainment of non--renormalizable higher--than--four canonical dimension operators by integrating out the heavy modes in KK theories has been sporadically discussed in the literature~\cite{RW}. Moreover, a remarkable feature of KK theories is the presence of massive gauge bosons, which must be subjected to a gauge--fixing procedure in order to be properly quantized. The possibility of having gauge dependent coefficients multiplying the non--renormalizable operators produced by the integration of heavy gauge KK modes comes into play. This interesting behavior, not discussed in the literature so far, is rather natural, as one--loop off--shell Green's functions involving gauge fields into the loops are not~\cite{NTtl,NTAIPp} necessarily, but often gauge dependent. The functional integration of heavy fields and the consequent derivation of a low--energy effective Lagrangian expansion is, in general, not an easy task. Even in the simplest cases, such as the Euler--Heisenberg Lagrangian~\cite{EHL}, the derivation of the low--energy expansion by integrating out the heavy fields (in such circumstances, the electron field) is intricate~\cite{DGMP}. Furthermore, the sole consideration of a Yang--Mills Lagrangian instead of the electromagnetic theory renders the obtainment of the corresponding effective Lagrangian quite a technical challenge. People has developed methods~\cite{inthfref} to calculate effective Lagrangian expansions by integrating out heavy fields. We follow the elegant approach given in Ref.~\cite{BS} and suitably adjust it to work in the case in which the heavy fields to integrate out are gauge fields. We wish to emphasize that the integration of heavy gauge fields, leading to an effective Lagrangian expansion, has not been done before. We first perform the calculation in the Feynman--t 'Hooft gauge, which renders the procedure the simplest it can be. After that, we consider the general case, in which the gauge--fixing parameter remains unfixed, and compare the resulting expression with that obtained in the Feynman--'t Hooft gauge. In both cases we obtain a low--energy effective Lagrangian expansion that incorporates up to SGT--invariant canonical--dimension--six operators.

The quantum Lagrangian, ${\cal L}_{\cal Q}$, can be divided into three parts as
\begin{equation}
{\cal L}_{\cal Q}={\cal L}_{\rm YM}+{\cal L}^{\rm 1-loop}_\xi+{\cal L}^{\rm heavy},
\label{llh}
\end{equation}
with ${\cal L}_{\rm YM}$ standing for the ordinary four dimensional Yang--Mills Lagrangian, which is purely constituted by zero--mode gauge fields. The ${\cal L}^{\rm 1-loop}_\xi$ term is a bridge that links the low--energy physics with the higher dimensional effects, and contains all of the one--loop corrections to light Green's functions. The last part of Eq.(\ref{llh}) comprehends contributions of KK excited modes that impact low--energy Green's functions, for the first time, at the two--loop level. As we are interested in the one--loop contributions to light Green's functions, from here on we disregard the ${\cal L}^{\rm heavy}$ term. The explicit form of ${\cal L}^{\rm 1-loop}_\xi$ is
\begin{eqnarray}
{\cal L}_\xi^{\rm 1-loop}&=&\frac{1}{2}g_{\mu\nu}\hspace{0.05cm}A^{(m)b\mu}\hspace{0.05cm}{\cal D}^{ba}_\alpha\hspace{0.05cm}{\cal D}^{ad\alpha}A^{(m)d\nu}+gf^{bad}A^{(m)b\mu}F^a_{\mu\nu}A^{(m)d\nu}-\frac{1}{2}\left( 1-\frac{1}{\xi} \right)A^{(m)b\mu}\hspace{0.05cm}{\cal D}^{ba}_\mu\hspace{0.05cm}{\cal D}^{ad}_\nu A^{(m)d\nu}
\nonumber  \\ \nonumber \\ &&
+\frac{1}{2}m_m^2\hspace{0.05cm}g_{\mu\nu}A^{(m)a\mu}A^{(m)a\nu}-\frac{1}{2}A^{(m)b}_5\hspace{0.05cm}{\cal D}^{ba}_\alpha\hspace{0.05cm}{\cal D}^{ad\alpha}A^{(m)d}_5-\frac{1}{2}\xi m_m^2A^{(m)a}_5A^{(m)a}_5
\nonumber \\ \nonumber \\ &&
+\hspace{0.05cm}\bar{C}^{(m)b}\hspace{0.05cm}{\cal D}^{ba}_\alpha\hspace{0.05cm}{\cal D}^{ad\alpha}C^{(m)d}-\xi m_m^2\bar{C}^{(m)a}C^{(m)a},
\end{eqnarray}
whose structure carry a latent gauge dependence through the gauge--fixing parameter. As we pretend to obtain a low--energy effective theory, we shall integrate out not only the heavy gauge degrees of freedom, but we shall also include the ghost and the pseudo--Goldstone bosons fields, so that we define the effective action, $S_{\rm eff}$, by
\begin{equation}
{\rm exp}\{ iS_{\rm eff} \}=\int {\cal D}A^{(n)}_\mu\hspace{0.05cm}{\cal D} A^{(n)}_5\hspace{0.05cm}{\cal D}\bar{C}^{(n)}\hspace{0.05cm}{\cal D}C^{(n)}{\rm exp}\{ iS_{\cal Q} \}= \int {\cal D}A^{(n)}_\mu\hspace{0.05cm}{\cal D}A^{(n)}_5\hspace{0.05cm}{\cal D}\bar{C}^{(n)}\hspace{0.05cm}{\cal D}C^{(n)}{\rm exp}\left\{ i\int d^4x\hspace{0.05cm}{\cal L}_{\cal Q} \right\},
\end{equation}
in which the functional integration affects all of the heavy KK modes of the theory. By integrating out the heavy KK modes, we obtain the low--energy effective action,
\begin{eqnarray}
S_{\rm eff}&=&S_{\rm YM}+\frac{i}{2}\sum_{m=1}^\infty{\rm Tr}\hspace{0.05cm}{\rm log}\left[ g_{\mu\nu}({\cal D}^2+m_m^2)-\left( 1-\frac{1}{\xi} \right){\cal D}_\mu{\cal D}_\nu-4igF_{\mu\nu} \right]
\nonumber \\ \nonumber \\&&
+\frac{i}{2}\sum_{m=1}^\infty{\rm Tr}\hspace{0.05cm}{\rm log}\left[ -{\cal D}^2-\xi m_m^2 \right]-i\sum_{m=1}^\infty{\rm Tr}\hspace{0.05cm}{\rm log}\left[ -2{\cal D}^2-2\xi m_m^2 \right],
\label{Seff1}
\end{eqnarray}
where $F_{\mu\nu}=F^a_{\mu\nu}T^a$, with $T^a$ representing the generators of the ${\rm SU}_4(N)$ gauge group. Besides, we denote ${\cal D}^2={\cal D}^\alpha{\cal D}_\alpha$, while the symbol ``Tr" indicates a trace over both the internal and the external degrees of freedom. The former of such degrees of freedom are the four dimensional space--time points, which are labeled by continuous indices, whereas the internal degrees of freedom are those corresponding to the ${\rm SU}_4(N)$ and four dimensional Lorentz groups. In this expression, $S_{\rm YM}$ is the standard four dimensional Yang--Mills action, which is defined by the ${\cal L}_{\rm YM}$ Lagrangian. All other terms contain the one--loop effects of heavy KK modes on light Green's functions. Note that these terms include an infinite sum that runs over all of the KK heavy modes. The first of the one--loop terms is produced by the heavy gauge KK modes and is the only one that includes a trace over Lorentz indices. Its structure greatly simplifies when taking the Feynman--'t Hooft gauge, that is, by electing $\xi=1$. The SGT--covariant gauge--fixing approach followed in the present work leads to notable simplifications through the cancelation of the unphysical couplings $A^{(m)a}_\mu A^{(n)b}_5$ and $A^{(0)a}_\mu A^{(m)b}_\nu A^{(n)c}_5$. As the gauge--fixing functions, Eq.(\ref{gffunc}), eliminate such terms, the contributions of the gauge and scalar fields are separated of each other, and this dissociation not only makes the derivation of the effective action, Eq.(\ref{Seff1}), easier, but also reveals an interesting relation among the contributions of the pseudo--Goldstone bosons and those of the ghost fields. Specifically, the second and third one--loop terms of the effective action come, respectively, from the pseudo--Goldstone bosons and ghost contributions. Note that the traces in such terms are essentially equal because the factor "2" appearing in the argument of the logarithm in the ghost contribution can be dropped, as it only contributes trivially to the effective action. This explicitly shows that the ghost fields contributions are minus twice times those of the pseudo--Goldstone bosons, which is an interesting feature that characterizes gauge--fixing procedures like the one followed here. Such a remarkable property has been fully exploited in KK theories~\cite{FMNRT} and also in other contexts~\cite{331tM,RTT}, for it simplifies loop calcuations involving both pseudo--Goldstone bosons and ghost contributions.

\subsection{The Feynman--'t Hooft gauge}
By taking the Feynman--'t Hooft gauge ($\xi=1$), the effective action, Eq.(\ref{Seff1}), reads
\begin{eqnarray}
S_{\rm eff}&=&S_{\rm YM}+\frac{i}{2}\sum_{m=1}^\infty{\rm Tr}\hspace{0.05cm}{\rm log}\left[ g_{\mu\nu}({\cal D}^2+m_m^2)-4igF_{\mu\nu} \right]-\frac{i}{2}\sum_{m=1}^\infty{\rm Tr}\hspace{0.05cm}{\rm log}\left[ -{\cal D}^2-m_m^2 \right].
\label{SeffFt}
\end{eqnarray}
The pure--gauge one--loop trace is the one most simplified, as under this particular gauge the term containing the cross derivatives ${\cal D}_\mu{\cal D}_\nu$ vanishes. The second part looks like the trace generated by integrating out a scalar field and has been already solved in Ref.{\cite{BS}}. For this calculation, we utilize the dimensional regularization scheme, so that, from here on, we work in $d$ dimensions. Under such circumstances, the term involving the metric tensor in the pure--gauge trace contributes $d$ times the scalar trace. In other words, one should expect
\begin{equation}
{\rm Tr}\hspace{0.05cm}{\rm log}\left[ g_{\mu\nu}({\cal D}^2+m_m^2) \right]=d\hspace{0.05cm}{\rm Tr}\hspace{0.05cm}{\rm log}\left[ -{\cal D}^2-m_m^2 \right]
\label{geqDs}
\end{equation}
to be fulfilled. Such an asseveration reads clearer if one thinks that the pure--gauge trace in Eq.(\ref{geqDs}) is generated by a functional integration of gauge fields, which are arranged as $d$--component vectors. Each of such components contributes as a scalar, leading to a total of $d$ scalar--like contributions. However, the leading contributions of the effective action, Eq.(\ref{SeffFt}), come from the second term in the pure--gauge trace. In Appendix \ref{appA}, we explicitly show how to obtain a low--energy expansion from a gauge trace like that in the first one--loop term of Eq.(\ref{SeffFt}). According to Eq.(\ref{finalloopexpansion}), we derive the following formula:
\begin{eqnarray}
i\hspace{0.05cm}{\rm Tr}\hspace{0.05cm}{\rm log}\left[ g_{\mu\nu}({\cal D}^2+M^2)+U_{\mu\nu}(x) \right]&=&\int d^dx\left[ \frac{1}{(4\pi)^2}M^2\left( \Delta_\epsilon+{\rm log}\left( \frac{\mu^2}{M^2} \right)+1 \right){\rm tr}\left\{ U^\mu\hspace{0.01cm}_\mu \right\}
\right.
\nonumber \\ \nonumber \\&&
+\frac{1}{(4\pi)^2}\frac{1}{2}\left( \Delta_\epsilon+{\rm log}\left( \frac{\mu^2}{M^2} \right) \right){\rm tr}\left\{ U_{\mu\nu}U^{\mu\nu} \right\}
\nonumber \\ \nonumber \\&&
\left.
-\frac{g^2}{(4\pi)^2}\frac{1}{3}\left( \Delta_\epsilon+{\rm log}\left( \frac{\mu^2}{M^2}\right)-\frac{1}{2}\right){\rm tr}\left\{ F_{\mu\nu}F^{\mu\nu} \right\}
\right]
\nonumber \\ \nonumber \\&& +\int d^4x\left[ -\frac{1}{(4\pi)^2}\frac{1}{6}\frac{1}{M^2}{\rm tr}\left\{ U_{\mu\nu}U^{\nu\sigma}U_\sigma\hspace{0.01cm}^\mu \right\}
+\frac{1}{(4\pi)^2}\frac{1}{3}\frac{1}{M^2}{\rm tr}\left\{ {\cal D}_\mu U^{\mu\nu}{\cal D}^\sigma U_{\sigma\nu} \right\}
\right.
\nonumber \\ \nonumber \\&&
-\frac{g^2}{(4\pi)^2}\frac{1}{3}\frac{1}{M^2}{\rm tr}\left\{ F_{\mu\nu}U^{\nu\sigma}F_\sigma\hspace{0.01cm}^\mu \right\}-\frac{g^2}{(4\pi)^2}\frac{1}{15}\frac{1}{M^2}{\rm tr}\left\{ {\cal D}_\mu F^{\mu\nu}{\cal D}^\sigma F_{\sigma\nu} \right\}
\nonumber \\ \nonumber \\&& \left.
-\frac{ig^3}{(4\pi)^2}\frac{2}{45}\frac{1}{M^2}{\rm tr}\left\{ F_{\mu\nu}F^{\nu\sigma}F_\sigma\hspace{0.01cm}^\nu \right\}
\right]+{\cal O}(1/M^4),
\label{FtgTr}
\end{eqnarray}
with
\begin{equation}
\begin{array}{lcr}
\displaystyle
\Delta_\epsilon=\frac{1}{\epsilon}-\gamma_E+{\rm log}(4\pi), && \displaystyle \epsilon=\frac{4-d}{2},
\end{array}
\end{equation}
and $U_{\mu\nu}(x)$ representing an arbitrary matrix--valued function of the space--time coordinates. In order to achieve this result, the traces of the external degrees of freedom and Lorentz indices have been taken, so that the symbol ``tr" denotes a trace only with respect to the gauge group. This formula, which gives a low--energy expansion up to canonical--dimension--six operators, holds for any gauge trace with such an structure, and can be employed in different contexts other than extra dimensions. On the other hand, from Ref.\cite{BS}, the expression
\begin{eqnarray}
i\hspace{0.05cm}{\rm Tr}\hspace{0.05cm}{\rm log}\left[ -({\cal D}^2+M^2) \right]&=&\int d^dx\left[ -\frac{g^2}{(4\pi)^2}\frac{1}{12}\left( \Delta_\epsilon+{\rm log}\left( \frac{\mu^2}{M^2} \right) \right){\rm tr}\left\{ F_{\mu\nu}F^{\mu\nu} \right\} \right]
\nonumber \\ \nonumber \\&&
+\int d^4x\left[ -\frac{ig^3}{(4\pi)^2}\frac{1}{90}\frac{1}{M^2}\hspace{0.05cm}{\rm tr}\left\{ F_{\mu\nu}F^{\nu\sigma}F_\sigma\hspace{0.001cm}^\mu \right\}
\right.
\nonumber \\ \nonumber \\&&
\left.
-\frac{g^2}{(4\pi)^2}\frac{1}{60}\frac{1}{M^2}{\rm tr}\left\{ {\cal D}_\mu F^{\mu\nu}{\cal D}^\sigma F_{\sigma\nu} \right\} \right]+{\cal O}(1/M^4)
\label{FtsTr}
\end{eqnarray}
is fulfilled. By setting $U_{\mu\nu}=-4igF_{\mu\nu}$ in Eq.(\ref{FtgTr}) and using the resulting expression altogether with Eq.(\ref{FtsTr}), the effective action, Eq.(\ref{SeffFt}), generates a low--energy effective theory,
\begin{eqnarray}
{\cal L}^1_{\rm eff}&=&{\cal L}_{\rm YM}+\frac{g^2}{(4\pi)^2}\frac{31}{8}\sum_{m=1}^\infty\left[ \Delta_\epsilon+{\rm log}\left( \frac{\mu^2}{m_m^2} \right)+\frac{2}{93} \right]{\rm tr}\left\{ F_{\mu\nu}F^{\mu\nu} \right\}-\frac{ig^3}{(4\pi)^2}\frac{281}{60}\sum_{m=1}^\infty\frac{R^2}{m^2}\hspace{0.05cm}{\rm tr}\left\{ F_{\mu\nu}F^{\nu\sigma}F_\sigma\hspace{0.001cm}^\mu \right\}
\nonumber \\ \nonumber \\&&
-\frac{g^2}{(4\pi)^2}\frac{323}{120}\sum_{m=1}^\infty\frac{R^2}{m^2}\hspace{0.05cm}{\rm tr}\left\{ {\cal D}_\mu F^{\mu\nu}{\cal D}^\sigma F_{\sigma\nu} \right\}+{\cal O}(R^4).
\label{Leff1exp}
\end{eqnarray}
The ${\cal L}^1_{\rm eff}$ effective Lagrangian has some interesting features that are worthy of commenting. This low--energy expansion includes the four dimensional ${\rm SU}_4(N)$--invariant Yang--Mills Lagrangian, ${\cal L}_{\rm YM}$, whose structure involves exclusively gauge KK zero modes, which are the light fields of the KK theory. This term, which is insensitive to the KK heavy modes, is renormalizable. The second term of the effective Lagrangian has UV divergencies and also discrete infinities generated by the KK sums. Nonetheless, it has no physical consequences, as it can be absorbed by the parameters of the low-energy theory. The effective Lagrangian also comprehends an infinite sum of higher--than--four canonical dimension operators that are supressed by appropriate powers of the compactification scale, $R^{-1}$. We have determined all of the terms that involve non--renormalizable canonical--dimension--six operators, which are governed by the ${\rm SU}_4(N)$ gauge group as well as the Lorentz group. Such operators belong to the list~\cite{nrinv} of canonical--dimension--six operators that are allowed by the symmetries of the four dimensional Yang--Mills theory. It is a remarkable quality of the non--renormalizable terms that the KK sums appearing in them are convergent. On the other hand, note that these terms do not contain UV divergencies. These asseverations are notable, as they explicitly prove that the one--loop effects of the KK excited modes on the light Green's functions are renormalizable, as it was recently shown~\cite{NT} by following a different approach. As the canonical--dimension--four part, ${\cal L}_{\rm YM}$, is renormalizable, the effective expansion is expected~\cite{W}, from the beginning, to fulfill the decoupling theorem~\cite{AC}. This decoupling behavior can be beheld in Eq.(\ref{Leff1exp}), where all of the effects of the extra dimensional physics trivially vanish in the limit of a very large (small) compactification scale (radius).

The philosophy of the effective Lagrangians formalism is simple, indeed. One can parametrize the effects of physics governing nature at a higher energy scale by "disguising" such effects as terms involving non--renormalizable operators that respect the low--energy continuous symmetries, although violations of the $C$, $P$, $T$  discrete transformations can arise. The low--energy parametrization occurs through the coefficients multiplying these non--renormalizable invariant operators, and such a description has a limited range of validity, below certain cut--off scale. As one gets closer to the energy scale characterizing the heavy physics, the operator expansion becomes senseless, and beyond the cut--off the dynamic variables and symmetries of the fundamental theory fully describe nature. The parameters carrying the heavy physics information can be determined in terms of the fundamental constants of the heavy physics theory, if such physical description is known beforehand. When we ignore the details of the high energy description, the effective Lagrangian parameters can be estimated or bounded by using experimental data. In such context, the effective Lagrangian formalism is a practical tool that allows one to study effects of heavy physics in a model--independent manner. In fact, the invariant ${\rm tr}\{ F_{\mu\nu}F^{\nu\sigma}F_\sigma\hspace{0.001cm}^\mu \}$ has been widely studied in the literature~\cite{Ow, NRST}, as it contributes to the static electromagnetic properties of the $W$ gauge boson. The non--renormalizable canonical--dimension--six operators of Eq.(\ref{Leff1exp}) produce, among other effects, CP--even contributions to the $WW\gamma$ and $WWZ$ vertices. An interesting issue concerning such effects is that their contributions to any multiloop Green's function are gauge independent with respect to the gauge fixing of the light fields. This property has been employed, for instance, to calculate~\cite{NRST} the one--loop contribution of the ${\rm tr}\{ F_{\mu\nu}F^{\nu\sigma}F_\sigma\hspace{0.001cm}^\mu \}$ operator to the neutrino charge radius. The coefficients of the low--energy effective Lagrangian expansion obtained in this section are written in terms of the compactification radius, which is the additional parameter incorporated by the extra dimension. The non--renormalizable terms in the ${\cal L}^1_{\rm eff}$ effective Lagrangian are phenomenologically interesting as they possess valuable information about the one--loop contributions of the heavy KK modes to low--energy Green's functions. For instance, in the context of the electroweak SM, one can calculate the KK excited modes one--loop contributions to the $S$, $T$, $U$ parameters~\cite{STU} from the ${\rm tr}\{ {\cal D}_\mu F^{\mu\nu}{\cal D}^\sigma F_{\sigma\nu} \}$ invariant through tree--level diagrams and find that all of such contributions vanish.

So far, we have worked in the context defined by the Feynman--'t Hooft gauge, which is convenient from the practical viewpoint and physically interesting. The quantization of gauge systems is engaging as the profound issues of gauge invariance and gauge independence play a central role. While gauge symmetry is one of the essential blocks that constitute any gauge theory, the quantization of such systems, however, requires the degeneracy associated with the gauge invariance to be removed, and the concept of gauge independence enters as a crucial feature when calculating physical observables. As gauge invariance ensures that any election of the gauge invariably leads to the same physical results, the S--matrix elements must be gauge independent, that is, they cannot depend on the gauge--fixing procedure. Contrastingly, it often occurs that the Green's functions are gauge dependent objects, atlhough such dependence must vanish when all of the Green's functions that contribute to an S--matrix element are taken together. Nevertheless, there exist the possibility of preserving some sort of gauge invariance through unconventional quantization schemes, such as the Background Field Method~\cite{BFM}(BFM) and the Pinch Technique~\cite{PT}(PT). Within the former approach, each of the gauge fields, ${\cal G}^a_\mu$, is splitted into a classical background field, $G^a_\mu$, and a quantum fluctuation, ${\cal Q}^a_\mu$: ${\cal G}^a_\mu\rightarrow G^a_\mu+{\cal Q}^a_\mu$. One then quantizes the quantum fields, for which a gauge--fixing procedure for them must be performed. The resulting quantum theory is still gauge invariant when the classical background fields are gauge--transformed, but it depends on a gauge--fixing parameter, $\xi_{\cal Q}$, originated in the fixation of the gauge for the quantum fluctuations. This procedure leads to Green's functions satisfying simple (QED--like) Ward identities, but involving the gauge--fixing parameter. Though this unconventional quantization approach produces gauge dependent Green's functions, it is expected to provide us with information quite close to the physical reality. The PT, on the other hand, is a diagrammatic method that pursues the construction of a quantum action that leads to both gauge invariant and gauge independent Green's functions. This method consists in constructing well--behaved Green's functions of a given number of points by combining some individual contributions from Green's functions of equal and higher numbers of points, whose Feynman rules are derived from a conventional effective action or even from a nonconventional scheme. Remarkably, the BFM and the PT are subtly related, which is suggested by the fact that the Green's functions built of BFM Feynman rules coincide with those calculated by utilizing the PT when the calculations are performed in the $\xi_Q=1$ gauge. This interesting link was first established~\cite{BFMPT1} at the one loop level, then confirmed~\cite{BFMPT2} at the two--loop level, and more recently it was shown~\cite{BFMPT3} that it is fulfilled at any order of perturbation theory. The explanation of such a striking connection remains so far unknown, but it is worth emphasizing that the Feynman--'t Hooft gauge does not produce unphysical thresholds. In the context of KK theories, the corrections of the heavy KK modes to light Green's functions in the Feynman--'t Hooft gauge yield~\cite{FMNRT} well--behaved results and provide an estimation of the effects of the KK excited fields. In this gauge, the coefficients multiplying the non--renormalizable canonical--dimension--six operators in the ${\cal L}^1_{\rm eff}$ effective Lagrangian are appraised to be of order $10^{-7}{\rm GeV}^2$ for a compactification scale as large as $R^{-1}=300$GeV~\cite{UED1}. Larger compactification scales shall lead to more stringent suppressions, as it happens, for instance, in the case of $R^{-1}=4$TeV~\cite{RW}, for which these coefficients are of order $10^{-9}{\rm GeV}^2$.

\subsection{The general $R_\xi$ gauge}
As it was shown above, the Feynman--'t Hooft gauge supplies an estimate of the suppression of the non--renormalizable canonical--dimension--six invariants in the low--energy effective Lagrangian. However, the possibility of obtaining a gauge dependent low--energy expansion is another issue that deserves attention. For that reason, in this section we work in the context of the $R_\xi$ gauge. In these more general circumstances, the gauge trace in the effective action, Eq.(\ref{Seff1}), is given by
\begin{equation}
i\hspace{0.05cm}{\rm Tr}\hspace{0.05cm}{\rm log}\left[ g_{\mu\nu}({\cal D}^2+m_m^2)-\left( 1-\frac{1}{\xi} \right){\cal D}_\mu{\cal D}_\nu-4igF_{\mu\nu} \right],
\end{equation}
where the cross derivative term remains. The presence of such a term makes the obtainment of the low--energy expansion quite complicated if one attempts to literally follow the method outlined in Appendix \ref{appA}. Nevertheless, the procedure can be greatly simplified by noting that four dimensional Lorentz covariance allows one to write the cross derivative term as
\begin{equation}
{\cal D}_\mu{\cal D}_\nu =\frac{1}{d}\hspace{0.05cm}g_{\mu\nu}{\cal D}^2-\frac{ig}{2}F_{\mu\nu},
\end{equation}
where the factor $1/d$ comes from the fact that we are working in $d$ dimensions. With this in mind, the gauge trace can be expressed as
\begin{equation}
\begin{array}{l}
\displaystyle i\hspace{0.05cm}{\rm Tr}\hspace{0.05cm}{\rm log}\left[ g_{\mu\nu}({\cal D}^2+m_m^2)-\left( 1-\frac{1}{\xi} \right){\cal D}_\mu{\cal D}_\nu-4igF_{\mu\nu} \right]
\\ \\ \displaystyle
=i\hspace{0.05cm}{\rm Tr}\hspace{0.05cm}{\rm log}\left[ g_{\mu\nu}\left( {\cal D}^2+\left( 1-\frac{\alpha}{d} \right)m_m^2 \right)-ig\left( 1-\frac{\alpha}{d} \right)\left( \frac{8-\alpha}{2} \right)F_{\mu\nu} \right],
\end{array}
\end{equation}
with $\alpha$ defined as
\begin{equation}
\alpha\equiv 1-
\frac{1}{\xi}.
\end{equation}
Written in this form, the gauge trace has the same structure as that of Eq.(\ref{FtgTr}), and employing the method of Appendix \ref{appA} is now profitable. On the other hand, the trace carrying the contributions of the pseudo--Goldstone bosons and the ghost fields looks, in this gauge, like the scalar trace in Eq.(\ref{FtsTr}), which can be utilized to obtain the corresponding expansion. By employing these results, we find the low--energy expansion to be
\begin{eqnarray}
{\cal L}^\xi_{\rm eff}&=&{\cal L}_{\rm YM}+\frac{g^2}{(4\pi)^2}\frac{7\alpha^2-120\alpha+496}{8(4-\alpha)^2}\sum_{m=1}^\infty\left( \Delta_\epsilon+{\rm log}\left( \frac{\mu^2}{m_m^2} \right)-\frac{4}{3}\frac{5\alpha^2-88\alpha+368}{7\alpha^2-120\alpha+496}{\rm log}\left( \frac{4}{4-\alpha} \right)
\right.
\nonumber \\ \nonumber \\&&
\left.
+\frac{(4-\alpha)^2}{3(7\alpha^2-120\alpha+496)}{\rm log}(1-\alpha)+\frac{2}{3}\frac{11\alpha^3-180\alpha^2+720\alpha+64}{(4-\alpha)(7\alpha^2-120\alpha+496)}
 \right){\rm tr}\left\{ F_{\mu\nu}F^{\mu\nu} \right\}
\nonumber \\ \nonumber \\&&
+\frac{ig^3}{(4\pi)^2}\frac{5\alpha^3-161\alpha^2+1528\alpha-4496}{60(4-\alpha)^2}\sum_{m=1}^\infty\frac{R^2}{m^2}\hspace{0.05cm}{\rm tr}\left\{ F_{\mu\nu}F^{\nu\sigma}F_\sigma\hspace{0.001cm}^\mu \right\}
\nonumber \\ \nonumber \\&& -\frac{g^2}{(4\pi)^2}\frac{20\alpha^2-323\alpha+1292}{120(4-\alpha)}\sum_{m=1}^\infty\frac{R^2}{m^2}\hspace{0.05cm}{\rm tr}\left\{ {\cal D}_\mu F^{\mu\nu}{\cal D}^\sigma F_{\sigma\nu} \right\} +{\cal O}(R^4).
\label{Lxeff}
\end{eqnarray}
This result is clearly gauge dependent, as the gauge--fixing parameter is embedded all over the coefficients of the non--renormalizable invariants. Note that, by virtue of the definition of $\alpha$, it occurs that $\alpha\rightarrow 0$ as $\xi\rightarrow 1$, that is, when taking the Feynman--'t Hooft gauge. In such case, we consistently obtain the ${\cal L}^1_{\rm eff}$ Lagrangian, which was derived in the last subsection by taking this particular gauge from the beginning. As commented above, the gauge--dependence of the low--energy effective expansion is not surprising. In general, the standard derivation~\cite{DGMP} of non--renormalizable terms by the integration of heavy fields in a given theory leads, at a first stage, to a local series of operators in which the low--energy symmetries are hidden. Each term of this series corresponds ~\cite{DGMP} to a one--loop diagram whose external lines are light fields, while its loops are constituted by the heavy fields that have been integrated out. As all particles in such terms (diagrams) are off--shell, the result of integrating out gauge fields that have been subjected to a gauge--fixing procedure naturally should lead to a gauge--dependent low--energy expansion, as it occurred in Eq.(\ref{Lxeff}). The term involving the invariant ${\rm tr}\{ F_{\mu\nu}F^{\mu\nu} \}$ in Eq.(\ref{Lxeff}) concentrates all of the UV divergencies, that is, those associated with the continuous as well as the discrete sums. However, these effects are unobservable, as they can be absorbed by the parameters of the light theory. On the other hand, the discrete sums appearing in the non--renormalizable invariants are finite and can be performed. As commented in the Introduction, the five dimensional Yang--Mills theory is not fundamental, and the impact of the physics beyond this extra dimensional theory can be parametrized by means of higher--than--five--canonical--dimension operators whose building blocks are the five dimensional dynamic variables and symmetries. When the compactification of the extra dimension is performed, these operators are expressed in terms of the KK modes and become invariant under the SGT and the NSGT. They are suppressed by powers of the cut--off $M_{\rm S}$, which defines the limit of validity of the five dimensional theory. Such a supression renders the effective effects of the fundamental description very weak, in particular those which involve KK excited modes. The less--suppressed terms of such operators are those constituted exclusively by KK zero modes. By considering only such less--suppressed terms altogether with the low--energy effective expansion exhibited in Eq.(\ref{Lxeff}), the resulting effective theory is composed exclusively by KK zero modes. In other words, the four dimensional effective theory so obtained involves only effects from the physics at the two heavy scales on low--energy Green's functions. Such Lagrangian is given by
\begin{eqnarray}
{\cal L}_{\rm eff}^{\xi}&=&{\cal L}_{\rm YM}+\frac{ig^3R^2}{5760}\hspace{0.05cm}\frac{5\alpha^3-161\alpha^2+1528\alpha-4496}{(4-\alpha)^2}\hspace{0.05cm}{\rm tr}\left\{ F_{\mu\nu}F^{\nu\sigma}F_\sigma\hspace{0.001cm}^\mu \right\}
\nonumber \\ \nonumber \\&& -\frac{g^2R^2}{11520}\hspace{0.05cm}\frac{20\alpha^2-323\alpha+1292}{(4-\alpha)}\hspace{0.05cm}{\rm tr}\left\{ {\cal D}_\mu F^{\mu\nu}{\cal D}^\sigma F_{\sigma\nu} \right\} +{\cal O}(R^4)+\sum_{k>4}\frac{\alpha_k}{M^{k-4}_{\rm S}}\hspace{0.05cm}{\cal O}^{\rm 4D}_k( A^{(0)a}_\mu ),
\label{finaleffL}
\end{eqnarray}
where the KK sums have been solved and the unobservable terms, absorbable by renormalization, have been omitted. The fourth term of the right--hand side of the last expression represents all higher--than--six-canonical--dimension non--renormalizable operators the emerge from integrating out the heavy KK fields. Of course, all of such SGT--invariants shall be suppressed by the compactification scale, $R^{-1}$, in powers that are equal or greater than 4. We have just indicated, through the last term in the right--hand side, the presence of effects from the physics beyond the extra dimensional description. This term contains all of the operators that parametrize physics at the $M_{\rm S}$ scale.  The precise expressions of such operators should be determined by KK--expanding, at the five dimensional level, the covariant objects composing the higher--than--five--canonical--dimension operators, which can be done once the compactification of the extra dimension has been performed. As established before, we are thinking only in those terms that involve zero KK modes, but no KK heavy fields, and this issue has been explicitly indicated in Eq.(\ref{finaleffL}). The $\alpha_k$ are dimensionless parameters that quantify the impact of the $M_{\rm S}$--scale physics at the four dimensional level. As these terms are suppressed by the fundamental energy scale, they are expected to be dominated by the non--renormalizable terms that we obtained in this paper. The fact has been phenomenologically illustrated~\cite{FMNRT}, recently.

It is worth comparing the impact of the physics beyond the fundamental scale with the contributions of the non--renormalizable terms that we derived by integrating out the KK excitations. As we discussed above, the Feynman--'t Hooft gauge is a physically suitable option to perform estimations concerning gauge dependent quantities. We take this gauge, so that the canonical--dimension--six terms of Eq.(\ref{finaleffL}) read
\begin{eqnarray}
\displaystyle
ig^3\alpha_{F^3} R^2{\rm tr}\{ F_{\mu\nu}F^{\nu\sigma}F_\sigma\hspace{0.001cm}^\mu \},
\label{effED1}
\\ \nonumber \\
\displaystyle
g^2\alpha_{DF}R^2{\rm tr}\{ {\cal D}_\mu F^{\mu\nu}{\cal D}^\sigma F_\sigma\hspace{0.001cm}^\nu \},
\label{effED2}
\end{eqnarray}
where the coefficients $\alpha_{F^3}$ and $\alpha_{DF}$ are both of order $10^{-2}$. On the other hand, the $M_{\rm S}$--scale physics could generate the invariants
\begin{eqnarray}
\int_0^{2\pi R}dy
\left(
\beta_{F^3}\frac{ig_5^3}{M_{\rm S}}{\rm tr}\{ {\cal F}_{MN}{\cal F}^{NS}{\cal F}_S\hspace{0.001cm}^M \}
\right)&=&\beta^{(0)}_{F^3} \frac{ig^3}{(M_{\rm S}R^{-1})}{\rm tr}\{ F_{\mu\nu}F^{\nu\sigma}F_\sigma\hspace{0.001cm}^\mu \}+\ldots ,
\label{Ms1}
\\ \nonumber \\
\int_0^{2\pi R}dy
\left(
\beta_{DF}\frac{g_5^2}{M_{\rm S}}{\rm tr}\{ D_M{\cal F}^{MN}D^{S}{\cal F}_{SN} \}
\right)&=&\beta^{(0)}_{DF}\frac{g^2}{(M_{\rm S}R^{-1})}{\rm tr}\{ {\cal D}_\mu F^{\mu\nu}{\cal D}^\sigma F_{\sigma\nu} \}+\ldots ,
\label{Ms2}
\end{eqnarray}
with $D_M$ standing for the ${\rm SU}_5(N)$ covariant derivative in the adjoint representation of the group, and where the dimensionless coefficients $\beta^{(0)}_{F^3}=2\pi \beta_{F^3}$ and $\beta^{(0)}_{DF}=2\pi\beta_{DF}$ parametrize, at low energy, the physics beyond $M_{\rm S}$. Note that the non--renormalizable operators explicitly shown in the right--hand side of these expressions are supressed with respect to those of Eqs.(\ref{effED1}) and (\ref{effED2}) by $R/M_{\rm S}$. The extra dimensional theory becomes strongly coupled at some scale, which could be identified with $M_{\rm S}$. In the context of the SM defined in a space--time with one UED, the requirement that the theory remains perturbative up to $M_{\rm S}$ has lead to the estimation~\cite{UED1,UED2} $M_{\rm S}\sim 30\hspace{0.01cm}R^{-1}$. In such circumstances, we take $\beta^{(0)}_{F^3}\sim \alpha_{F^3}$ and $\beta^{(0)}_{DF}\sim \alpha_{DF}$, in Eqs.(\ref{Ms1}) and (\ref{Ms2}), and use them to determine that the contributions from the fundamental physics are about 3\% of those from the extra dimensional physics. 

\section{Conclusions}
In this paper we have analyzed some theoretical aspects about a pure--gauge Kaluza--Klein theory originated in the compactification of the fifth dimension in the context of a ${\rm SU}_5(N)$--invariant Yang--Mills theory defined in a five--dimensional space--time manifold. The dynamic variables of this KK theory are the KK modes, which can be divided into two types: the zero KK modes, which are the light fields, and which coincide with the ordinary four dimensional Yang--Mills gauge fields; and the KK excited modes, which are manifestations of the fifth dimensional theory and are the heavy fields. The GKKT is invariant under two sorts of gauge transformations that we call the standard gauge transformations and the non--standard gauge transformations. It is possible to fix the gauge with respect to the KK heavy modes and so remove the degeneracy associated to the NSGT, but leaving that with respect to the SGT. We have taken advantage of this interesting issue to derive a low--energy effective Lagrangian expansion. Such an effective theory has been defined by the functional integration over all of the KK excited modes, which comprehend gauge fields as well as pseudo--Golsdtone bosons and ghost fields. We have explicitly proven that our gauge--fixing scheme renders the one--loop contributions of the ghost fields minus twice those produced by the pseudo--Goldstone bosons. This interesting relation is a property of non--conventional quantization aproaches like the one followed in this paper. By integrating out the heavy fields, we have obtained an effective Lagrangian that involves up to canonical--dimension--six non--renormalizable SGT--invariant operators, built of light fields, exclusively. In order to achieve such an expression, we have followed and appropriately adjusted an elegant method, proposed in the literature some years ago, to derive low--energy expansions. The functional integration of heavy gauge fields to obtain a low--energy effective expansion is a novel calculation of the present paper. Our result involves all of the independent canonical--dimension--six invariants whose building blocks are the four dimensional Yang--Mills fields. We utilized this expansion to compare the effects of the extra dimensional Yang--Mills theory on light physics with those originated in the fundamental description of nature beyond the cut--off scale $M_{\rm S}$, and found that the impact of the latter is negligible with respect to the significance of the former. We have first calculated this low--energy effective Lagrangian in the Feynman--'t Hooft gauge, and have found that all of the divergencies, of both discrete and continuous origins, can be absorbed by the parameters of the low--energy theory, which implies that the one--loop contributions of the KK excited modes to light Green's functions are renormalizable, as it was proven reciently in the literature from a different perspective. The KK sums in the effective Lagrangian are all convergent, so that the non--renormalizable character of the five--dimensional theory does not manifest itself at the one--loop level. This asseveration is not necessarily true for two--loops or higher order calculations, or when two or more extra dimensions are considered. The fact that the infinite KK sums incarnate a latent source of divergencies in calculations involving two--loop or higher order contributions of KK heavy fields to light Green's functions is an awkward feature of the theory. Nevertheless, the first and dominant corrections to the light physics enter at the one--loop level as renormalizable effects. This quality allows one to obtain unambiguous results, which endows the theory with predictive power. We have also derived the effective Lagrangian expansion in the general $R_\xi$ gauge, and have found that the low--energy expansion is gauge dependent with respect to the fixation of the gauge for the KK excited modes. This feature is consistent with the fact that the one--loop Green's functions containing gauge fields
into the loops exhibit gauge dependence through the presence of the gauge--fixing parameter in their structure. This is a physically crucial issue that has not been discussed before in the literature.

\acknowledgments{ We acknowledge financial support from CONACYT and SNI (M\' exico).}

\appendix

\section{The solution for a gauge determinant}
\label{appA}
The functional integration of heavy fields in a theory describing physics at certain high energy scale leads to determinants that concern all of the light degrees of freedom, both internal and external. The determinants so obtained can be transformed into traces over all of such degrees of freedom. In this appendix, we adjust the method presented in Ref.~\cite{BS} to calculate a trace carrying the contributions of heavy gauge bosons and obtain a low--energy expansion up to canonical--dimension--six non--renormalizable operators. Consider the general trace
\begin{equation}
i\hspace{0.05cm}{\rm Tr}\hspace{0.05cm}{\rm log}\left[ g_{\mu\nu}({\cal D}^2+M^2)+U_{\mu\nu}(x) \right]\equiv\int d^4x\hspace{0.05cm}{\cal L}_{\rm 1-loop}(x)
\label{Trdef}
\end{equation}
where ${\cal D}_\mu$ is the covariant derivative for the ${\rm SU}(N)$ gauge group and $U_{\mu\nu}(x)$ is a space--time dependent matrix that we suppose to be arbitrary. The covariant derivative is given by
\begin{equation}
\begin{array}{lr}
{\cal D}_\mu=\partial_\mu+G_\mu,&\hspace{0.5cm}G_\mu=-igT^aG^a_\mu,
\end{array}
\end{equation}
with $T^a$ representing the generators of the gauge group and $G^a_\mu$ standing for the gauge fields. The curvature, which we denote by $G_{\mu\nu}^a$, is defined in terms of the covariant derivative as
\begin{equation}
\begin{array}{lr}
G_{\mu\nu}=\left[ {\cal D}_\mu,{\cal D}_\nu \right],&\hspace{0.5cm}G_{\mu\nu}=-igT^aG^a_{\mu\nu}.
\label{curvadef}
\end{array}
\end{equation}
The trace operation in Eq.(\ref{Trdef}) acts on the points of the space--time, which are the external degrees of freedom. It also affects the internal degrees of freedom, which in this case are determined by the gauge and Lorentz groups. In the following, the symbol ``Tr" shall refer to a trace over both the external and the internal degrees of freedom, while ``tr" shall indicate a trace over internal degrees of freedom, exclusively. As divergencies shall appear, below, they must be appropriately regularized. We follow the dimensional regularization approach, for which we work, from here on, in $d$ dimensions. Note that the argument of the trace is non--local, so that, up to this point, performing this operation makes no sense. To obtain a local expression, one can first perform the trace over the space--time coordinates, which, for a general operator ${\cal O}$, should be understood as
\begin{equation}
{\rm Tr}\left\{ {\cal O} \right\}=\int d^dx\hspace{0.1cm}{\rm tr}\hspace{0.05cm}\langle x|{\cal O}|x\rangle=\int d^dx\hspace{0.05cm}d^d\tilde{p}\hspace{0.1cm}{\rm tr}\left\{ \langle x | {\cal O} | p \rangle \langle p | x \rangle \right\}
\end{equation}
where a completeness relation has been inserted and we have defined
\begin{equation}
d^d\tilde{p}=\mu^{(4-d)/2}\frac{d^dp}{(2\pi)^d},
\end{equation}
so that $\mu$ is a factor introduced to appropriately correct dimensions. For a general quantum state, $|\alpha\rangle$,
\begin{equation}
\langle x | {\cal O} | \alpha\rangle={\cal O}_x\langle x|\alpha\rangle={\cal O}_x\hspace{0.05cm}\alpha(x),
\end{equation}
with ${\cal O}_x$ standing for the operator ${\cal O}$ in the representation of positions. With this in mind, note that
\begin{equation}
{\rm Tr}\{ {\cal O} \}=\int d^dx\hspace{0.05cm}d^d\tilde{p}\hspace{0.1cm}{\rm tr}\left\{ e^{ip\cdot x}{\cal O}_xe^{-ip\cdot x} \right\}
\label{Tredf}.
\end{equation}
By defining $\Pi_\mu\equiv i{\cal D}_\mu$ and then applying the general result shown in Eq.(\ref{Tredf}), along with the operator identity $e^{ip\cdot x}f(\Pi)e^{-ip\cdot x}=f(\Pi+p)$, to the gauge trace, Eq.(\ref{Trdef}), one obtains
\begin{eqnarray}
i\hspace{0.05cm}{\rm Tr}\hspace{0.05cm}{\rm log}\left[ g_{\mu\nu}({\cal D}^2+M^2)+U_{\mu\nu} \right]&=&i\hspace{0.05cm}{\rm Tr}\hspace{0.05cm}{\rm log}\left[ g_{\mu\nu}\left( -\Pi^2+M^2 \right)+U_{\mu\nu} \right]
\nonumber \\ \nonumber \\&=& i\int d^dx\hspace{0.05cm}d^d\tilde{p}\hspace{0.1cm}{\rm tr}\left\{ {\rm log}\left[ -p^2+M^2 \right]+{\rm log}\left[ g_{\mu\nu}+\frac{(\Pi^2+2\Pi\cdot p)g_{\mu\nu}-U_{\mu\nu}}{p^2-M^2} \right] \right\}{\bf 1}
\end{eqnarray}
In this expression, the logarithm operators act on the identity, which we denote by ``{\bf 1}". The first term of the argument of the trace in the second line of the last expression contains no fields, and hence contributes only to the vacuum energy density. Thus, we drop it in what follows and conserve only the second term, so that the gauge trace is expanded as
\begin{equation}
i\hspace{0.05cm}{\rm Tr}\hspace{0.05cm}{\rm log}\left[ g_{\mu\nu}({\cal D}^2+M^2)+U_{\mu\nu} \right]=i\int d^dx\hspace{0.05cm}d^d\tilde{p}\hspace{0.1cm}{\rm tr}\sum_{k=1}^\infty\frac{(-1)^{k+1}}{k}\frac{\left[ (\Pi^2+2\Pi\cdot p)\delta^\mu\hspace{0.001cm}_\nu-U^\mu\hspace{0.001cm}_\nu \right]^k}{\left[ p^2-M^2 \right]^k}\hspace{0.05cm}{\bf 1}.
\label{Trexpansion}
\end{equation}
The argument of the gauge trace, written in this form, is local, so that the trace over the internal degrees of freedom can be taken in each term of the series.

From the local form of Eq.(\ref{Trexpansion}), one can appreciate that the calculation of the momentum integrals shall provide an expansion of non--renormalizable operators, each one multiplied by a power of $M$. In other words, by comparing Eqs.(\ref{Trdef}) and (\ref{Trexpansion}), one can extract the Lagrangian
\begin{equation}
{\cal L}_{\rm 1-loop}=\sum_{k=1}^\infty\frac{(-1)^{k+1}}{k}\hspace{0.05cm}{\rm tr}\left\{ i\int d^d\tilde{p}\frac{\left[ (\Pi^2+2\Pi\cdot p)\delta^\mu\hspace{0.001cm}_\nu-U^\mu\hspace{0.001cm}_\nu \right]^k}{\left[ p^2-M^2 \right]^k} {\bf 1} \right\},
\label{appL1loopexp}
\end{equation}
then solve the loop integrals term by term in the series, and finally write ${\cal L}_{\rm 1-loop}$ as
\begin{equation}
{\cal L}_{\rm 1-loop}=d\sum_{k=1}^\infty\frac{c_k}{M^{2k-4}}{\cal O}_k,
\label{Lexpgenform}
\end{equation}
with ${\cal O}_k$ representing a linear combination of traces of gauge invariant operators of canonical dimension $2k$, built of the gauge fields $G^a_\mu$, the matrix $U_{\mu\nu}$, and the ${\rm SU}(N)$ covariant derivative. In the ${\cal L}_{\rm 1-loop}$ expansion shown in Eq.(\ref{Lexpgenform}), there is a global factor, $d$, which is expected because the gauge trace, Eq.(\ref{Trdef}), was produced by the integration of vector fields, which are constituted by $d$ scalar fields, and the sum of all contributions produces this global factor. A convenient normalization of this expansion fixes the $c_k$ coefficients as
\begin{equation}
c_k=\frac{1}{(4\pi)^2}\left( \frac{M^2}{4\pi\mu^2} \right)^{d/2-2}\Gamma\left( k-\frac{d}{2} \right).
\label{normalization}
\end{equation}
As already commented, the ${\cal O}_k$ are combinations of traces of dimension--2$k$ operators, so that, in general, they should have the form
\begin{equation}
{\cal O}_k=\sum_j a_{k,j}\hspace{0.05cm}{\cal O}_{k,j}.
\label{tracescombination}
\end{equation}
For the first values of $k$ ($=1,2,3$), we shall employ the following sets as bases:
\begin{eqnarray}
k=1:&\hspace{0.5cm}&\left({\rm tr}\{ U^\mu\hspace{0.001cm}_\mu \}\right)
\label{bk1}
\\ \nonumber \\
k=2:&\hspace{0.5cm}&\left( {\rm tr}\{ U_{\mu\nu} U^{\nu\mu} \},d\hspace{0.05cm}{\rm tr}\{ G_{\mu\nu}G^{\mu\nu} \} \right)
\label{bk2}
\\ \nonumber \\
k=3:&\hspace{0.5cm}&\left( {\rm tr}\{ U_{\mu\nu}U^{\nu\sigma}U_\sigma\hspace{0.001cm}^\mu \},{\rm tr}\{ {\cal D}_\mu U^{\mu\nu}{\cal D}^\sigma U_{\sigma\nu} \},{\rm tr}\{ G_{\mu\nu}U^{\nu\sigma}G_\sigma\hspace{0.001cm}^\mu \},d\hspace{0.05cm}{\rm tr}\{ {\cal D}_\mu G^{\mu\nu}{\cal D}^\sigma G_{\sigma\nu} \},d\hspace{0.05cm}{\rm tr}\{ G_{\mu\nu}G^{\nu\sigma}G_\sigma\hspace{0.001cm}^\mu \} \right)
\label{bk3}
\end{eqnarray}
The method proposed in Ref.~\cite{BS} relies on the fact that Eqs.(\ref{appL1loopexp}), (\ref{Lexpgenform}) and (\ref{tracescombination}) are valid for any field configuration. In fact, the $a_{k,j}$ coefficients are just numbers, independent of the field configuration, so that if one is able to determine them within a specific choice, the coefficients corresponding to the bases exhibited in Eqs.(\ref{bk1}), (\ref{bk2}) and (\ref{bk3}) can be obtained.  An appropriate election is the field configuration such that
\begin{equation}
\begin{array}{lll}
E_\mu\equiv G_\mu,&\hspace{0.1cm}\partial_\mu E_\nu=0,&\hspace{0.01cm}U^\mu\hspace{0.001cm}_\nu=-\delta^\mu\hspace{0.001cm}_\nu E^2,
\label{scconditions1}
\end{array}
\end{equation}
which imply that
\begin{equation}
\begin{array}{ll}
G_{\mu\nu}=[E_\mu,E_\nu],&\hspace{0.1cm}{\cal D}_\mu{\cal M}=[E_\mu,{\cal M}],
\label{scconditions2}
\end{array}
\end{equation}
where ${\cal M}$ represents any matrix valued function of $E_\mu$ and $U^\mu\hspace{0.001cm}_\nu$. In these very particular circumstances, the numerator of the ${\cal L}_{\rm 1-loop}$ expansion, Eq.(\ref{appL1loopexp}), is greatly simplified, for each term of the series can be expressed as
\begin{equation}
\left[ (\Pi^2+2\Pi\cdot p)\delta^\mu\hspace{0.001cm}_\nu -U^\mu\hspace{0.001cm}_\nu \right]^k{\bf 1}=\delta^\mu\hspace{0.001cm}_\nu(2iE\cdot p)^k,
\end{equation}
which considerably reduces the procedure of calculating the momentum integrals:
\begin{eqnarray}
{\cal L}_{\rm 1-loop}(E_\mu)&=&\sum_{k=1}^\infty\frac{(-1)^{k+1}}{k}\hspace{0.05cm}{\rm tr}\left\{ \int d^d\tilde{p}\hspace{0.05cm}\frac{\delta^\mu\hspace{0.001cm}_\nu (2iE\cdot p)^k}{(p^2-M^2)^k} \right\}
\nonumber \\ \nonumber \\
&=&d\hspace{0.05cm}\sum_{k=1}^\infty\frac{(-1)^{k+1}4^k}{2k}\hspace{0.05cm}i\int d^d\tilde{p}\hspace{0.05cm}\frac{p^{\mu_1}p^{\mu_2}\ldots p^{\mu_{2k}}}{(p^2-M^2)^{2k}}\hspace{0.05cm}{\rm tr}\left\{ E_{\mu_1}\ldots E_{\mu_{2k}} \right\}
\nonumber \\ \nonumber \\
&=&d\hspace{0.05cm}\sum_{k=1}^\infty\frac{1}{M^{2k-4}}\frac{1}{(4\pi)^2}\left( \frac{M^2}{4\pi\mu^2} \right)^{d/2-2}\Gamma\left( k-\frac{d}{2} \right)\frac{2^k}{(2k)!}\hspace{0.05cm}{\rm tr}\left\{ {\cal S}_{2k}(E_\mu) \right\}
\nonumber \\ \nonumber \\
&=&d\hspace{0.05cm}\sum_{k=1}^\infty\frac{c_k}{M^{2k-4}}\frac{2^k}{(2k)!}\hspace{0.05cm}{\rm tr}\left\{ {\cal S}_{2k}(E_\mu) \right\} ,
\label{steps}
\end{eqnarray}
In passing from the first to the second line in Eq.(\ref{steps}), we took the trace over the Lorentz indices, which gave rise to the $d$ factor (recall that we are working in $d$ dimensions!). We then considered the fact that any loop integral with an odd number of momentum factors vanishes. From the second to the third line, we solved the momentum integrals by utilizing the result
\begin{equation}
i\int d^d\tilde{p}\hspace{0.1cm}\frac{p^{\mu_1}p^{\mu_2}\cdots p^{\mu_{2k}}}{(p^2-M^2)^{2k}}=(-1)^{k+1}\left( \frac{M^2}{4\pi\mu^2} \right)^{d/2-2}\frac{1}{(4\pi)^2}\frac{1}{M^{2k-4}}\frac{\Gamma(k-d/2)}{2^k\Gamma(2k)}S_{k}^{\mu_1 \mu_2\ldots\mu_{2k}}.
\end{equation}
In this expression, $S_{k}^{\mu_1\ldots\mu_{2k}}$ is a totally symmetric tensor built of the sum of all the products of $k$ metric tensors involving all the possible permutations of Lorentz indices. For instance, $S_2^{\mu_1\ldots\mu_4}=g^{\mu_1\mu_2}g^{\mu_3\mu_4}+g^{\mu_1\mu_3}g^{\mu_2\mu_4}+g^{\mu_1\mu_4}g^{\mu_2\mu_3}$. Also, we have employed the definition
\begin{equation}
{\cal S}_{2k}(E_\mu)\equiv S_k^{\mu_1\ldots\mu_{2k}}E_{\mu_1}\ldots E_{\mu_{2k}},
\end{equation}
so that ${\cal S}_{2k}(E_\mu)$ is the sum of all possible permutations of products of $2k$ $E_\mu$ fields in which all of such fields are Lorentz--contracted. For example, ${\cal S}_4(E_\mu)=(E^2)^2+E_\mu E_\nu E^\mu E^\nu+E_\mu E^2E^\mu$. Finally, from the third to the fourth line, we have used the normalization of the $c_k$, Eq.(\ref{normalization}). By comparing the last line of Eq.(\ref{steps}) with the general expansion exhibited in Eq.(\ref{Lexpgenform}), one can identify
\begin{equation}
{\cal O}^{\rm S}_k=\frac{2^k}{(2k)!}\hspace{0.05cm}{\rm tr}\{ {\cal S}_{2k}(E_\mu) \},
\label{expsc}
\end{equation}
with the superscript "S" indicating that we are working in the specific field configuration. Within this especial configuration, the ${\cal O}_k$, can be expanded as
\begin{equation}
{\cal O}^{\rm S}_k=\sum_ia^{\rm S}_{k,j}{\cal O}^{\rm S}_{k,j}.
\end{equation}
In this context, an appropriate set of bases of traces ${\cal O}^{\rm S}_{k,j}$ is
\begin{eqnarray}
k=1:&\hspace{0.5cm}&\left( {\rm tr}\{ E^2 \} \right)
\\ \nonumber \\
k=2:&\hspace{0.5cm}&\left( {\rm tr}\{ E_\mu E_\nu E^\mu E^\nu \},{\rm tr}\{ (E^2)^2 \} \right)
\\ \nonumber \\
k=3:&\hspace{0.5cm}&\left( {\rm tr}\{ E_\mu E_\nu E_\sigma E^\mu E^\nu E^\sigma \},{\rm tr}\{ (E^2)^3 \},{\rm tr}\{ E^2E_\mu E^2E^\mu \},{\rm tr}\{ E_\mu E_\nu E^\mu E_\sigma E^\nu E^\sigma \},{\rm tr}\{ E^2E_\mu E_\nu E^\mu E^\nu \} \right).
\end{eqnarray}
By employing Eq.(\ref{expsc}), one can straightforwardly obtain the ${\cal O}_k$ combinations in the especial configuration:
\begin{eqnarray}
{\cal O}^{\rm S}_1&=&{\rm tr}\{ E^2 \},
\label{sccombO1}
\\ \nonumber \\
{\cal O}^{\rm S}_2&=&\frac{1}{3}{\rm tr}\{ (E^2)^2 \}+\frac{1}{6}{\rm tr}\{ E_\mu E_\nu E^\mu E^\nu \},
\label{sccombO2}
\\ \nonumber \\
{\cal O}^{\rm S}_3&=&\frac{1}{90}{\rm tr}\{ E_\mu E_\nu E_\sigma E^\mu E^\nu E^\sigma \}+\frac{1}{45}{\rm tr}\{ (E^2)^3 \}+\frac{1}{30}{\rm tr}\{ E^2E_\mu E^2E^\mu \}+\frac{1}{30}{\rm tr}\{ E_\mu E_\nu E^\mu E_\sigma E^\nu E^\sigma \}
\nonumber \\ \nonumber \\
&&+\frac{1}{15}{\rm tr}\{ E^2E_\mu E_\nu E^\mu E^\nu \}.
\label{sccombO3}
\end{eqnarray}
One can also write the combinations ${\cal O}_k$ in terms of the bases in Eqs.(\ref{bk1}), (\ref{bk2}) and (\ref{bk3}), according to the general expression shown in Eq.(\ref{tracescombination}), and specialize the results to the special configuration, which was defined through Eqs.(\ref{scconditions1}) and (\ref{scconditions2}), as
\begin{equation}
{\cal O}^{\rm S}_k=\left.\sum_ia_{k,j}{\cal O}_{k,j}\right|_{\rm S}.
\end{equation}
So far, the $a_{k,j}$ coefficients remain unknown, but by equalizing the resulting expressions to Eqs.(\ref{sccombO1}), (\ref{sccombO2}) and (\ref{sccombO3}) as
\begin{equation}
{\cal O}^{\rm S}_k=\left.\sum_ia_{k,j}{\cal O}_{k,j}\right|_{\rm S}=\sum_ia_{k,j}^{\rm S}{\cal O}_{k,j}^{\rm S},
\end{equation}
one can determine such coefficients, which are independent of the configuration. We find
\begin{eqnarray}
{\cal O}_1&=&-\frac{1}{d}{\rm tr}\{  U^\mu\hspace{0.001cm}_\mu \},
\\ \nonumber \\
{\cal O}_2&=&\frac{1}{2d}{\rm tr}\{ U_{\mu\nu}U^{\nu\mu} \}+\frac{1}{12}{\rm tr}\{ G_{\mu\nu}G^{\mu\nu} \},
\\ \nonumber \\
{\cal O}_3&=&-\frac{1}{6d}{\rm tr}\{ U_{\mu\nu}U^{\nu\sigma}U_\sigma\hspace{0.001cm}^\mu \}+\frac{1}{12}{\rm tr}\{ {\cal D}_\mu U^{\mu\nu}{\cal D}^\sigma U_{\sigma\nu} \}+\frac{1}{12}{\rm tr}\{ G_{\mu\nu}U^{\nu\sigma}G_\sigma\hspace{0.001cm}^\mu \}+\frac{1}{60}{\rm tr}\{ {\cal D}_\mu G^{\mu\nu}{\cal D}^\sigma G_{\sigma\nu} \}
\nonumber \\ \nonumber \\
&&-\frac{1}{90}{\rm tr}\{ G_{\mu\nu}G^{\nu\sigma}G_\sigma\hspace{0.001cm}^\mu \}.
\end{eqnarray}
By inserting these results into Eq.(\ref{Lexpgenform}) along with the $c_k$ coefficients given by Eq.(\ref{normalization}), we obtain the following low--energy expansion,
\begin{eqnarray}
{\cal L}_{\rm 1-loop}&=&\frac{1}{(4\pi)^2}M^2\left( \Delta_\epsilon+{\rm log}\left( \frac{\mu^2}{M^2} \right)+1 \right){\rm tr}\{ U^\mu\hspace{0.001cm}_\mu \}+\frac{1}{(4\pi)^2}\frac{1}{2}\left( \Delta_\epsilon+{\rm log}\left( \frac{\mu^2}{M^2} \right) \right){\rm tr}\{ U_{\mu\nu}U^{\mu\nu} \}
\nonumber \\ \nonumber \\&&
+\frac{1}{(4\pi)^2}\frac{1}{3}\left( \Delta_\epsilon+{\rm log}\left( \frac{\mu^2}{M^2}-\frac{1}{2} \right) \right){\rm tr}\{ G_{\mu\nu}G^{\mu\nu} \}-\frac{1}{(4\pi)^2}\frac{1}{M^2}\frac{1}{6}{\rm tr}\{ U_{\mu\nu}U^{\nu\sigma}U_\sigma\hspace{0.001cm}^\mu \}
\nonumber \\ \nonumber \\&&
+\frac{1}{(4\pi)^2}\frac{1}{M^2}\frac{1}{3}{\rm tr}\{ {\cal D}_\mu U^{\mu\nu}{\cal D}^\sigma U_{\sigma\nu} \}+\frac{1}{(4\pi)^2}\frac{1}{M^2}\frac{1}{3}{\rm tr}\{ G_{\mu\nu}U^{\nu\sigma}G_\sigma\hspace{0.001cm}^\mu \}
\nonumber \\ \nonumber \\&&
+\frac{1}{(4\pi)^2}\frac{1}{M^2}\frac{1}{15}{\rm tr}\{ {\cal D}_\mu G^{\mu\nu}{\cal D}^\sigma G_{\sigma\nu} \}-\frac{1}{(4\pi)^2}\frac{1}{M^2}\frac{2}{45}{\rm tr}\{ G_{\mu\nu}G^{\nu\sigma}G_\sigma\hspace{0.001cm}^\mu \}.
\label{finalloopexpansion}
\end{eqnarray}
with
\begin{eqnarray}
\Delta_\epsilon=\frac{1}{\epsilon}-\gamma_E+{\rm log}(4\pi),&\hspace{0.5cm}&\epsilon=\frac{4-d}{2}.
\end{eqnarray}
As a final remark, note that, by virtue of Eq.(\ref{curvadef}), one should perform the change $G_{\mu\nu}\rightarrow-ig\hspace{0.01cm}G_{\mu\nu}$ in Eq.(\ref{finalloopexpansion}) in order to be in agreement with the standard notation.

\end{document}